\documentclass[preprint,showpacs,preprintnumbers,amsmath,amssymb, prb]{revtex4}
\usepackage{graphicx}
\usepackage{dcolumn}
\usepackage{bm}
\begin{document}

\title{Superzone gap formation and low lying crystal electric field levels in PrPd$_2$Ge$_2$ single crystal}

\author{Arvind Maurya}
\email{arvindmaurya.physics@gmail.com}
\affiliation{Department of Condensed Matter Physics and Materials
Science, Tata Institute of Fundamental Research, Homi Bhabha Road,
Colaba, Mumbai 400 005, India.}

\author{S. K. Dhar}
\affiliation{Department of Condensed Matter Physics and Materials
Science, Tata Institute of Fundamental Research, Homi Bhabha Road,
Colaba, Mumbai 400 005, India.}

\author{A. Thamizhavel}
\affiliation{Department of Condensed Matter Physics and Materials
Science, Tata Institute of Fundamental Research, Homi Bhabha Road,
Colaba, Mumbai 400 005, India.}

\date{\today}

\begin{abstract}
The magnetocrystalline anisotropy exhibited in PrPd$_2$Ge$_2$ single crystal has been investigated by measuring the magnetization, magnetic susceptibility, electrical resistivity and heat capacity.  PrPd$_2$Ge$_2$ crystallizes in the well known 
ThCr$_2$Si$_2$\--type tetragonal structure. The antiferromagnetic ordering  is confirmed as 5.1~K with the [001]-axis as the easy axis of magnetization.  A superzone gap formation is observed from the electrical resistivity measurement when the current is passed along the [001] direction.  The crystal electric field (CEF) analysis on the magnetic susceptibility, magnetization and the heat capacity measurements confirms a doublet ground state with a relatively low over all CEF level splitting.  The CEF level spacings and the Zeeman splitting at high fields become comparable and lead to metamagnetic transition at 34~T due to the CEF level crossing.

\end{abstract}



\maketitle

\section{Introduction}
Praseodymium (Pr$^{3+}$), being a non-Kramer's ion exhibits a variety of interesting magnetic behaviour in its compounds. Pr$^{3+}$ often orders magnetically with a doublet ground state, while it behaves as a Van Vleck paramagnet down to the lowest temperature when the crystal electric field split ground state is a singlet (e. g.~in PrRhAl$_4$Si$_2$~\cite{PrRhAl4Si2}). There is a surge in the study of Pr compounds in recent times after the observation of heavy fermion superconductivity in PrOs$_4$Sb$_{12}$, PrV$_2$Al$_{20}$, and  PrTi$_2$Al$_{20}$ at ambient or under pressure, originating from the quadrupolar Kondo order of Pr $f$\--orbitals~\cite{Bauer_PrOsSb, Sakai}.

PrPd$_2$Ge$_2$ is a member of the well known large family of compounds crystallizing in the tetragonal ThCr$_2$Si$_2$\--type structure (space group $I4/mmm$, \# 139). A previous report on polycrystalline sample provided evidence for an antiferromagnetic transition at 5~K~\cite{Welter}. Further, neutron diffraction indicated a magnetic cell three times larger than the chemical cell and the Pr moments aligned along the $c$\--axis with a  spontaneous magnetization of $\sim$2.0~$\mu_{\rm B}$ at 2~K~\cite{Welter}. Three possible configurations of the moments in the antiferromagnetic state, compatible with the tripling of the magnetic cell, were presented. However, the authors of that work felt that neutron diffraction on a single crystal was necessary to determine unambiguously the magnetic configuration. In the present work, we explore the magnetic properties of a single crystal of PrPd$_2$Ge$_2$, using the techniques of magnetization, electrical resistivity and heat capacity. Our data on single crystalline sample suggest that the easy direction of magnetization  may not lie exactly along the \textbf{c}-axis. The crystal electric field (CEF) analysis reveals an interesting possibility of metamagnetism at high fields ($\simeq$ 35~T) along the [001] direction due to the crossover among the  split energy levels with magnetic field. 

\section{Crystal growth}
We used the Czochralski  method to grow a single crystal of PrPd$_2$Ge$_2$ , in  a tetra-arc furnace, as the compound melts congruently. First, a homogeneous polycrystalline ingot, weighing  $\sim$10~gm was prepared by repeated arc melting of the  high pure metals in the stoichiometric ratio $1:2:2$, which was subsequently used as starting material for the crystal growth.  To start with, a tungsten rod was used as seed to pull the single crystal out of molten charge.  The pulling speed was maintained at 10~mm/hour, after the initial necking and stabilization. In order to estimate the magnetic part of the heat capacity and to estimate the magnetic entropy, a polycrystalline sample of the non-magnetic  LaPd$_2$Ge$_2$ was also prepared in a home built mono arc furnace.

To confirm the phase purity of grown crystal, a small portion of the specimen was subjected to powder x-ray diffraction (XRD).  The XRD (not shown here for brevity)  revealed  a clear pattern without any impurity peaks suggesting the single phase nature of the grown crystal.  A Le-bail fit was performed on the x-ray diffraction pattern and the lattice constants were estimated to be $a$~=~4.336(8)~\AA~ and $c$~=~10.050(8)~\AA,~ respectively, which are in good agreement with the previously reported values~\cite{Rossi,Welter}. The composition of the crystal was further confirmed by energy dispersive analysis by x-ray (EDAX). In order to study the anisotropic physical properties, the grown single crystal was cut along the principal crystallographic directions \textit{viz.}, [100] and [001] using a spark erosion cutting machine and back-reflection Laue diffraction.  Well defined Laue diffraction spots together with the four fold symmetry ascertain the good quality of the grown single crystal.

\section{Magnetic susceptibility and magnetization}
\begin{figure}[!]
\centering
\includegraphics[width=0.45\textwidth]{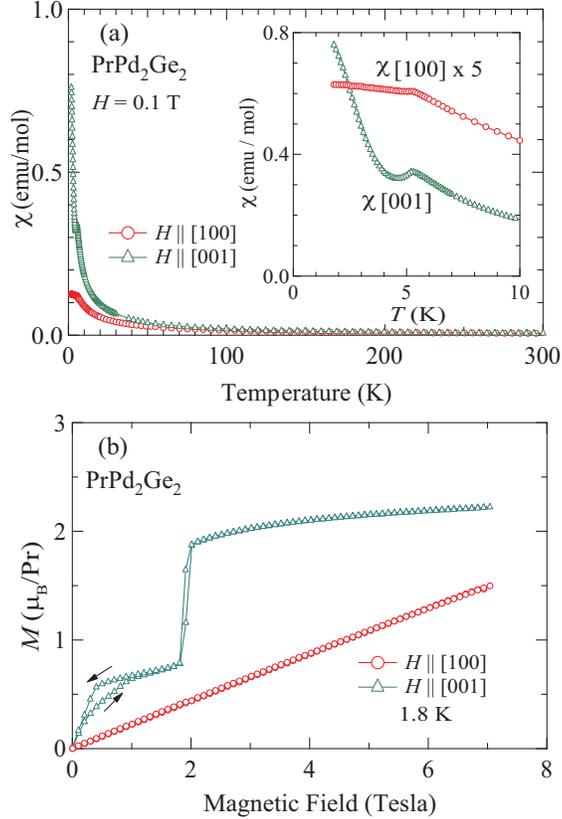}
\caption{\label{Pr_Magnetization}  (a) Magnetic susceptibility of PrPd$_2$Ge$_2$ between 1.9 and  300~K. Inset shows low temperature data on expanded scale, in which $\chi$ along [100] is multiplied by a factor of 5 to show its comparison with that of [001], (b) The isothermal magnetization data at 1.8~K along [100] and [001].}
\end{figure}

Magnetic susceptibility of PrPd$_2$Ge$_2$ shows a clear anomaly at $T_{\rm N}$ = 5.1~K,  with field parallel to [100] and [001] directions, respectively (see, Fig.\ref{Pr_Magnetization}), in conformity with the reported $T_{\rm N}$ of polycrystalline sample. Similar to isotypic CePd$_2$Ge$_2$~\cite{Arvind CePd2Ge2}, PrPd$_2$Ge$_2$ also exhibits a significant anisotropy in the magnetic susceptibility at low temperatures, measured in a field of 0.1~T. Magnitude of $\chi$ along [100] is smaller than that of [001], and remains nearly temperature independent in the antiferromagnetic state below $T_{\rm N}$, marking it as the hard axis of magnetization, while $\chi$ along [001] increases sharply at lower temperature after exhibiting a cusp at $T_{\rm N}$, unlike the conventional behaviour for a simple two sublattice antiferromagnet.  The increase in the magnetic susceptibility along the [001] direction at temperatures below   $T_{\rm N}$ reveals that the magnetic structure is complex as determined from the neutron diffraction data  by Welter and Halich~\cite{Welter}. 
%

The Curie-Weiss fit to the inverse susceptibility data gives $\mu_{\rm eff}$~=~3.70 (3.76) $\mu_{\rm B}$/Pr, and $\theta_{\rm p}$~=~-14.9 (4.3)~K in [100] and [001] directions, respectively. $\mu_{\rm eff}$ values are close to 3.58~$\mu_{\rm B}$, predicted by the Hund's rules for free Pr$^{3+}$. A negative value of $\theta_{\rm p}$ corresponds to antiferromagnetic correlations among Pr ions; however $\theta_{\rm p}$ has a positive, albeit small, value along [001], which is tentatively ascribed to ferromagnetic interaction among the next nearest neighbours in the (001) plane.

The isothermal magnetization at 1.8~K along [100] varies linearly with field, which is in conformity with the basal plane being the hard plane of magnetization. On the other hand the magnetization along the tetragonal [001] direction, undergoes a distinct  spin flip transition at 1.9 T. A change in the antiferromagnetic configuration presumably happens in the low field region ($\sim$1 T) as well which is marked by a distinct hysteresis. The magnetization attained at 7~T is 2.22 and 1.50 $\mu_{\rm B}$/Pr for field parallel to [001] and [100] directions, respectively. These values are significantly lower than the saturation value of 3.20~$\mu_{\rm B}$ for Pr$^{3+}$ ion corresponding to the total angular momentum quantum number $J=4$ and Land{\'e} g-factor ($g_J=4/5$). A higher magnetic field is required to populate all CEF split energy levels in order to attain the full moment value (see section~\ref{PrCEF}).  It may be noted that for a bipartite collinear antiferromagnet, the susceptibility along the easy axis gradually decreases to zero as the temperature is gradually decreased to zero. The isothermal magnetization should be nearly zero up to spin flop transition. We do not see such behavior in our single crystal. We believe our data show that the Pr moments do not lie parallel to [001] axis and the neutron diffraction results reported in Ref.~\onlinecite{Welter} are oversimplified. 
\section{Electrical Resistivity}
%
%
\begin{figure*}[!]
\centering
\includegraphics[width=0.85\textwidth]{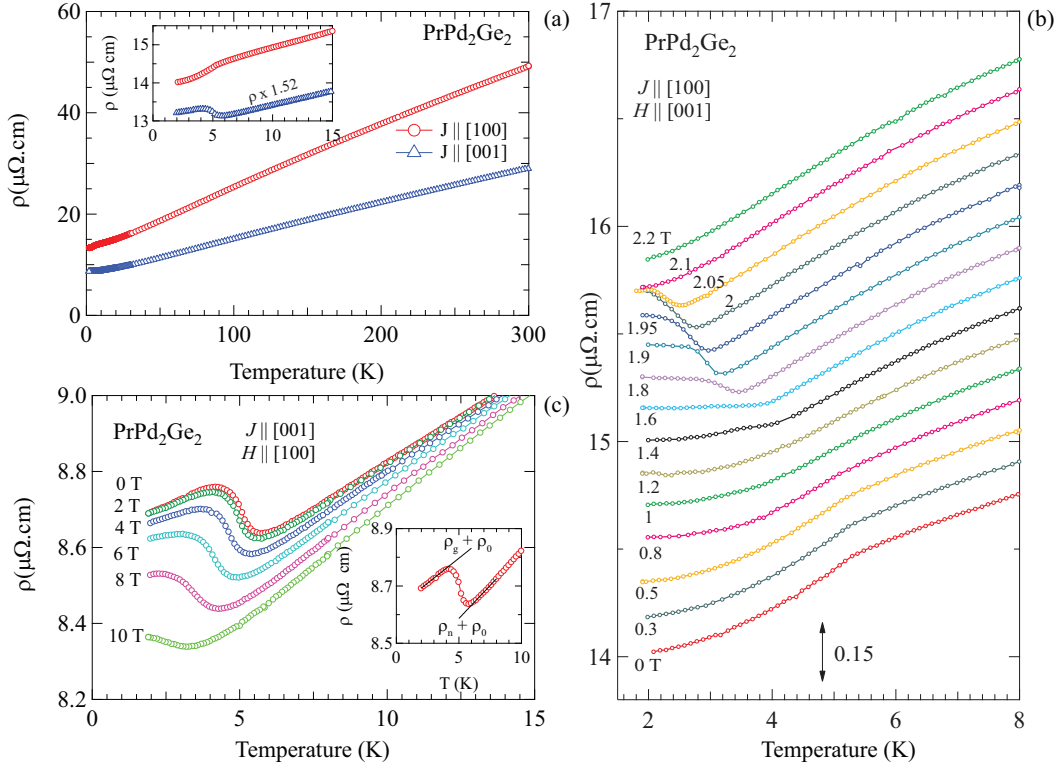}
\caption{\label{Pr_RT}  (a) Variation of zero field electrical resistivity $\rho$ with temperature, with current density $J$ parallel to principal crystallographic directions in PrPd$_2$Ge$_2$. Inset zooms up the low temperature part, for $J\parallel$~[001] direction the data is multiplied by 1.52 to show up in the same $y$-axis scale. (b) and (c) show the $\rho(T)$ data at selected transverse field in both configurations, respectively. Various iso-field traces of resistivity in (b) are shifted vertically for clarity.  The inset in (c) shows the low temperature electrical resistivity for $J\parallel$~[001], the solid lines are linear fit to the resistivity data to estimate the $\rho_n$ and $\rho_g$, the resistivities of the normal and the gapped state.}
\end{figure*}
From the electrical resistivity data, recorded between 2 and 300 K, shown in Fig.~\ref{Pr_RT}, PrPd$_2$Ge$_2$ is metallic down to 2~K. The resistivity is anisotropic and exhibits higher values for the current density  $J$ parallel to [100] direction. Note that below  $T_{\rm N}$, for $J\parallel$~[100], a faster drop resulting from the loss of magnetic disorder scattering is observed. But, for $J\parallel$~[001], $\rho(T)$ shows an upturn at $T_{\rm N}$. Such a feature observed in some antiferromagnets is generally attributed to a reduction in electron density caused by the opening of a gap (superzone gap) in the Fermi surface at $T_{\rm N}$  when the magnetic periodicity is incongruent with the periodicity of the chemical unit cell. Fig.~\ref{Pr_RT}(b) shows the data for $J\parallel$~[100] and $H\parallel$~[001]. It is noticed that the decrease of resistivity below $T_{\rm N}$ becomes less prominent as the field increases such that  at intermediate transverse field (1.40-2.05 T), a characteristic signature of superzone gap is observed (Fig.~\ref{Pr_RT}(b)). Whether the superzone gap persists to higher fields can be ascertained only by data taken at temperatures lower than 2~K. This interesting observation shows that in PrPd$_2$Ge$_2$, the superzone gap doesn't depend only on direction, but it is also a function of magnetic field. On the other hand, for $J\parallel$~[001] and $H\parallel$~[100], the zero field superzone gap persists at least up to 10 T (Fig.~\ref{Pr_RT}(c)). However, the   overall magnitude of resistivity, the upturn at $T_{\rm N}$  and $T_{\rm N}$ decrease. It is to be mentioned here that the jump in the electrical resistivity at the superzone gap in zero field is very small, roughly estimated to be 0.12~$\mu \Omega\cdot$cm indicates Fermi surface gap is very small in this case.  In order to estimate the magnitude of Fermi surface gapping, we followed the technique used by Mun et al.~\cite{Mun} for YbPtBi by calculating the relative change in the conductivities  using the relation:  ($\sigma_n - \sigma_g)/\sigma_n$, where $\sigma_n (= 1/\rho_n)$ and $\sigma_g (= 1/\rho_g)$ are the  conductivities of normal and the gapped states.  The conductivity values obtained below and above the superzone gap are shown in Table~\ref{Table1}.  It is obvious from the table that the Fermi surface gapping is only 3\% and it is almost constant for fields up to 4~T.  In order to get more insight into the behaviour of the Fermi surface gapping, experiments down to low temperature are necessary.

\begin{table}[]
\centering
\caption{The electrical resistivity values obtained from the linear fit (refer to inset of Fig.~\ref{Pr_RT}(b)) to the electrical resistivity data at temperatures below and above the superzone gap in different magnetic fields.   The last column gives the degree of Fermi surface gapping estimation.}
\label{Table1}
\begin{tabular}{c|c|c|c|c|c}
\hline
\begin{tabular}[c]{@{}c@{}}Magnets Field\\  (T)\end{tabular} & \begin{tabular}[c]{@{}c@{}}Resistivity\\ $\rho_g$\\ ($\mu \Omega \cdot $cm)\end{tabular} & \begin{tabular}[c]{@{}c@{}}Conductivity\\ $\sigma_g = 1/\rho_g$\\ ($\mu \Omega \cdot $cm)$^{-1}$\end{tabular} & \begin{tabular}[c]{@{}c@{}}Resistivity\\ $\rho_n$\\ ($\mu \Omega \cdot $cm)\end{tabular} & \begin{tabular}[c]{@{}c@{}}Conductivity\\ $\sigma_n = 1/\rho_n$\\ ($\mu \Omega \cdot $cm)$^{-1}$\end{tabular} & $\Delta \sigma = \frac{\sigma_n - \sigma_g}{\sigma_n}$ \\ \hline
0                                                            & 8.6229                                                                                   & 0.115970                                                                                                      & 8.3555                                                                                   & 0.11968                                                                                                       & 0.03099                                                \\ \hline
1.6                                                          & 8.6294                                                                                   & 0.11588                                                                                                       & 8.3600                                                                                   & 0.11962                                                                                                       & 0.03126                                                \\ \hline
2.0                                                          & 8.6260                                                                                   & 0.11593                                                                                                       & 8.3670                                                                                   & 0.11952                                                                                                       & 0.03004                                                \\ \hline
2.6                                                          & 8.6165                                                                                   & 0.11606                                                                                                       & 8.3558                                                                                   & 0.11968                                                                                                       & 0.03025                                                \\ \hline
4.0                                                          & 8.6189                                                                                   & 0.11602                                                                                                       & 8.3258                                                                                   & 0.12011                                                                                                       & 0.03405                                                \\ \hline
\end{tabular}
\end{table}
%
\section{Heat capacity}
\begin{figure}[!]
\centering
\includegraphics[width=0.45\textwidth]{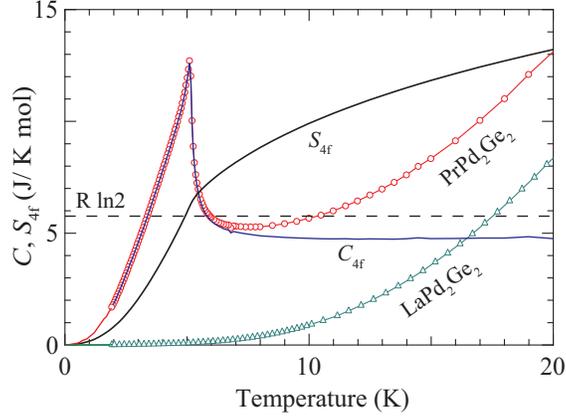}
\caption{\label{Pr_HC}  Heat capacity (C) as a function of temperature in PrPd$_2$Ge$_2$ and non magnetic analog LaPd$_2$Ge$_2$. The $4f$ contribution $C_{4f}$ and calculated entropy $S_{4f}$ are are also shown on the same scale. Dashed horizontal line marks the Rln2 value of entropy, where R is the gas constant.}
\end{figure}

The heat capacity of PrPd$_2$Ge$_2$ and non magnetic reference compound LaPd$_2$Ge$_2$ measured between 1.9~K and 20~K, is shown in Fig.~\ref{Pr_HC}.
The antiferromagnetic transition at 5.1~K is marked by a sharp peak. The low temperature anomaly observed on a polycrystalline sample of PrPd$_2$Ge$_2$ by Welter and Halich~\cite{Welter} at 3~K is not observed in our heat capacity data, indicating the good quality of our sample.  The heat capacity of non magnetic iso-structural LaPd$_2$Ge$_2$ was subtracted from that of PrPd$_2$Ge$_2$, to estimate the contribution by Pr-$4f$ electrons only ($C_{4f}$), assuming identical phonon contribution to the heat capacity in two compounds. Entropy  $S_{4f}$  was calculated by the same process as described for  CePd$_2$Ge$_2$. A smooth extrapolation of $C(T)$ data below 1.9~K in PrPd$_2$Ge$_2$ was done as shown by the solid line in Fig.~\ref{Pr_HC}, to get an approximate value of the heat capacity at low temperatures for which the experimental data are not available. The error in the estimation of entropy due to extrapolation is at the most a few \% of its true value.
%
%
 At $T_{\rm N}$, $S_{4f}$ is comparable to Rln2  corresponding to a doublet ground state with effective spin $S~=~1/2$. Here R is the universal gas constant. From the $C_{4f}$ values at higher temperatures, the CEF split levels have been derived, which is the subject of next section. 
\section{Crystal electric field  analysis}\label{PrCEF}
A crystal electric field (CEF) analysis was performed on the magnetization and heat capacity data of PrPd$_2$Ge$_2$ to derive possible information about the crystal electric field level splittings. For the case of Pr ($J=4, L=5, S=1$), CEF Hamiltonian can be expressed as,
\begin{equation}
\label{eqn_CEF_Hamiltonial}
\mathcal{H}_{{\rm CEF}}=(B_{2}^{0}O_{2}^{0}+B_{4}^{0}O_{4}^{0}+B_{4}^{4}O_{4}^{4}+  B_{6}^{0}O_{6}^{0} + B_{6}^{4}O_{6}^{4})+g\mu_{\rm B}\textbf{J}.\textbf{H},
\end{equation}
where $B_{l}^{m}$'s are CEF parameters and $O_{l}^{m}$'s are Steven's operators~\cite{Hutchings, Stevens}, respectively. Second term represents Zeeman energy in presence of magnetic field.

 Pr is a non Kramer's ion (integral spin $S$). In the tetragonal $\mathcal{D}_{\rm 4h}$ point symmetry, the CEF splits the 9 fold degenerate state of Pr$^{3+}$ ion into two doublets and 5 singlets~\cite{Runciman}. The compounds which exhibit singlet ground state are non magnetic (for instance the recently discovered PrRhAl$_4$Si$_2$)~\cite{PrRhAl4Si2}.
%
\begin{table*}[!]
\begin{center}
\caption{\label{table_CEF_Pr} CEF parameters, energy levels
and the corresponding wave functions for PrPd$_2$Ge$_2$ at zero field.}
\begin{tabular}{lccccccccc}\hline
\multicolumn{9}{c}{CEF parameters} \\ \hline
\multicolumn{2}{c}{$B_{2}^{0}$~(K)} & \multicolumn{2}{c}{$B_{4}^{0}$~(K)} & \multicolumn{2}{c}{$B_{4}^{4}$~(K)} & \multicolumn{2}{c}{$B_{6}^{0}$~(K)} & \multicolumn{2}{c}{$B_{6}^{4}$~(K)} \\
\multicolumn{2}{c}{$-1.1889$} & \multicolumn{2}{c}{$0.0147$} & \multicolumn{2}{c}{$0.0147$}	& \multicolumn{2}{c}{$3.9490\times10^{-4}$} & \multicolumn{2}{c}{$1.9220\times10^{-5}$} \\
\multicolumn{9}{c}{$\lambda_{[100]}$ = $0.1569$~mol/emu, 	$\lambda_{[001]}$ = $-1.28424 $~mol/emu} \\ \hline
\multicolumn{9}{c}{energy levels and wave functions}\\ \hline
$E$(K) & $\mid-4\rangle$ & $\mid-3\rangle$ & $\mid-2\rangle$ & $\mid-1\rangle$ & $\mid0\rangle$ & $\mid+1\rangle$& $\mid+2\rangle$ & $\mid+3\rangle$& $\mid+4\rangle$ \\
80.40	&	-0.2449	&	0	&	0	&	0	&	0.9381	&	0	&	0	&	0	&	-0.2450\\
80.23	&	0	&	0	&	0.7071	&	0	&	0	&	0	&	-0.7071	&	0	&	0\\
79.26	&	0	&	0	&	0	&	0.9507	&	0	&	0	&	0	&	-0.3101	&	0\\
79.26	&	0	&	-0.3101	&	0	&	0	&	0	&	0.9507	&	0	&	0	&	0 \\
27.17	&	0	&	0	&	0.7071	&	0	&	0	&	0	&	0.7071	&	0	&	0 \\
24.06	&	-0.7071	&	0	&	0	&	0	&	0	&	0	&	0	&	0	&	0.7071\\
16.37	&	0.6633	&	0	&	0	&	0	&	0.3464	&	0	&	0	&	0	&	0.6633\\
0	&	0	&	0.9507	&	0	&	0	&	0	&	0.3101	&	0	&	0	&	0\\
0	&	0	&	0	&	0	&	0.3100	&	0	&	0	&	0	&	0.9507	&	0 \\
\hline
\end{tabular}
\end{center}
\end{table*}
Since PrPd$_2$Ge$_2$ orders at 5.1~K, the ground state is definitely doublet, as indicated strongly by the entropy (see above). Therefore, the excited states comprise of a ground state doublet and 5 singlets and another doublet at some higher energy. 
\begin{figure*}[!]
\centering
\includegraphics[width=.85\textwidth]{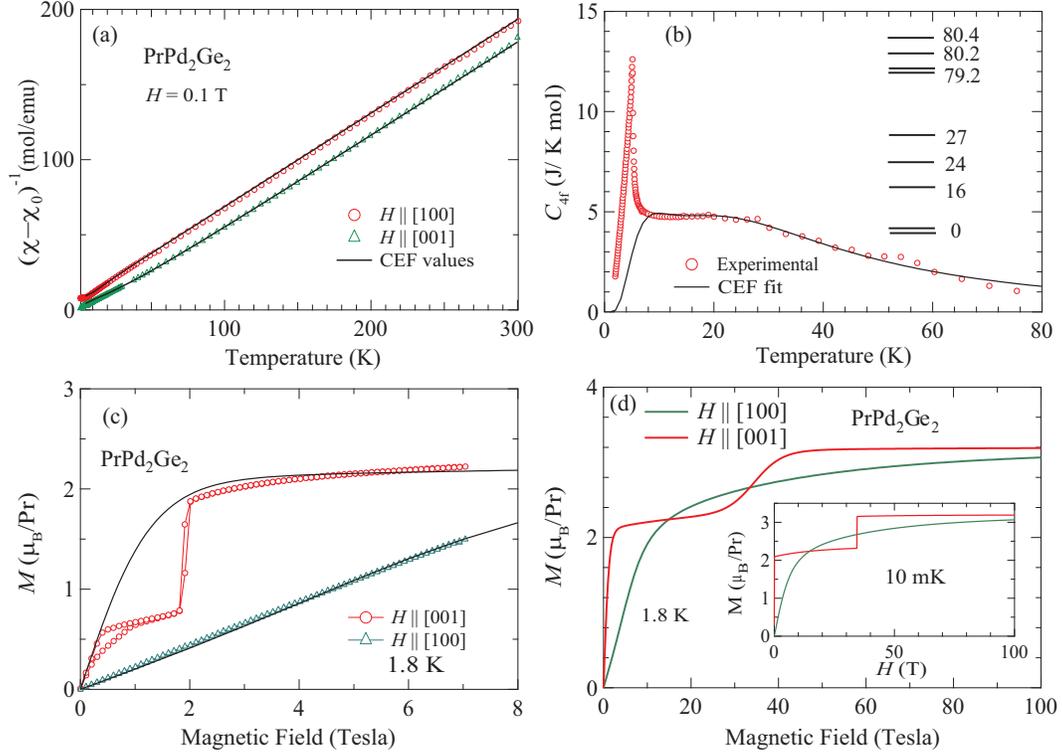}
\caption{\label{Pr_CEF} A comparison of calculated (a) inverse susceptibility at 0.1~T, (b) heat capacity and (c) isothermal magnetization derived from CEF eigenstates and eigen values with the corresponding experimental data. (d) Calculated magnetization of PrPd$_2$Ge$_2$ up to 100~Tesla at 1.8~K (main panel) and at 10~mK (inset) }
\end{figure*}
\begin{figure}[!]
\centering
\includegraphics[width=0.45\textwidth]{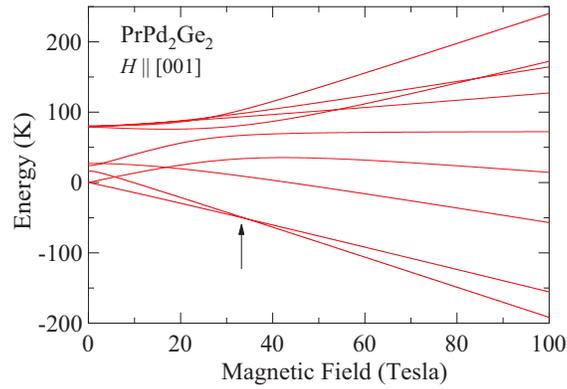}
\caption{\label{Pr_CEF2} Evolution of energy eigen values of CEF and Zeeman split Pr-energy levels with magnetic field along [001]\--direction in PrPd$_2$Ge$_2$. Arrow marks the ground state level crossing, which leads to metamagnetism (cf.~Fig.~\ref{Pr_CEF}(d)). }
\end{figure}

After diagonalizing the $(2J+1)\times (2J+1)$ i.e. 9$\times$9 dimensional matrices, the CEF parameters $B_l^m$, which fit the anisotropic susceptibility, isothermal magnetization (Fig.~\ref{Pr_CEF}c) and Schottky heat capacity of PrPd$_2$Ge$_2$, are listed in Table~\ref{table_CEF_Pr} and the computed curves are shown in Fig.~\ref{Pr_CEF} along with the experimental data. It may be noted that the heat capacity of both PrPd$_2$Ge$_2$ and LaPd$_2$Ge$_2$ was measured up to 80~K for this purpose. The ground state at zero magnetic field is an admixture of  $\mid-1\rangle$ and $\mid+3\rangle$ pure states.
There is a good degree of agreement between theory and observation, which gives credence to the correctness of the CEF parameters and energy levels derived from our analysis. However, advanced techniques like inelastic neutron scattering are required to confirm CEF splittings.

The main panel of Fig.~\ref{Pr_CEF}(d) shows the  calculated magnetization for applied field up to 100 T. The magnetic field removes the degeneracy, and their energy evolution with magnetic field is governed by Eq.~\ref{eqn_CEF_Hamiltonial}. The magnetization in [100] direction increases gradually with field and attains a marginally lower than full moment value of Pr (3.2 $\mu_{\rm B}$) at 100 T where it has not yet achieved saturation. Interestingly, along the [001] direction, there is a meta magnetic jump at around  34~Tesla and the magnetization attains the theoretical saturation value at higher fields. The metamagnetic jump at 34 T has a different origin than the spin-flop transition that occurs at ~ 1.9 T due to the change in the antiferromagnetic configuration with field. The metamagnetic jump at 34 T originates from the level crossing of the ground and the first excited CEF levels at that field (see Fig.~\ref{Pr_CEF2}) due to the dominance of Zeeman splitting over CEF splitting. When the magnetization is calculated for a temperature of 10 mK (inset of Fig.~\ref{Pr_CEF}(d)), the metamagnetic transition due to level crossings is nearly vertical as one would expect it to be in the ground state, free of effects due to thermal fluctuations at higher temperature. It would be interesting to measure the magnetization at higher fields to confirm the results of our CEF based calculations.
\section{Conclusion}
The anisotropic magnetic properties of PrPd$_2$Ge$_2$ have been investigated by growing a single crystal, in a tetra-arc furnace.  The phase purity and the crystal composition were confirmed by x-ray diffraction and EDAX.  The transport and magnetic properties reveal large anisotropy along the two principal crystallographic directions \textit{viz.}, [100] and [001].  The antiferromagnetic order is confirmed at $T_{\rm N}$ = 5.1~K. The increase in the magnetic susceptibility below $T_{\rm N}$ for $H~\parallel$~[001], the easy axis direction, clearly reveals the complex magnetic structure of this compound. The electrical resistivity data of PrPd$_2$Ge$_2$ show the existence  of anisotropic superzone gap which, interestingly, is dependent on both the crystallographic direction and applied magnetic field.  The N\'{e}el temperature was found to decrease with increasing field, as expected for a typical antiferromagnet system.   Our crystal field calculation explains  the anisotropy in the magnetic susceptibility and magnetization of PrPd$_2$Ge$_2$.  The energy levels determined from CEF analysis clearly explain the Schottky anomaly in the $4f$\--derived part of the heat capacity. The CEF levels in PrPd$_2$Ge$_2$ are comparatively low lying, which makes Zeeman interaction comparable to CEF splitting at a magnetic field of 34~Tesla leading to metamagnetic behaviour for $H\parallel$~[001]  due to level crossing. Its confirmation by high field magnetization on a single crystalline sample is desirable.


\begin{thebibliography}{99}

\bibitem{PrRhAl4Si2}A. Maurya, R. Kulkarni, A. Thamizhavel, and S. K. Dhar, Solid State Commun. {\bf 240}, 24 (2016)

\bibitem{Bauer_PrOsSb}E. Bauer, N. Frederick, P.-C. Ho, V. Zapf, and M. Maple, Phys. Rev. B {\bf 65}, 100506 (2002).

\bibitem{Sakai} A. Sakai, and S. Nakatsuji, J. Phys. Soc. Jpn. {\bf 80}, 063701 (2011).

\bibitem{Rossi}D. Rossi, R. Marazza, and R. Ferro, J. Less-Common Met. {\bf 66}, 17 (1979).

\bibitem{Welter} R. Welter, and K. Halich, J. Phys. Chem. Solids {\bf 67}, 862 (2006).

\bibitem{Arvind CePd2Ge2}A. Maurya, R. Kulkarni, S. K. Dhar, and A. Thamizhavel, J. Phys.: Condens. Matter {\bf 25}, 435603 (2013).

\bibitem{Mun}E. D. Mun, S. L. Bud'ko, C. Martin, H. kim, M. A. Tanatar, J. -H.Park, T. Murphy, G. M. Schmiedeshoff, N. Dilley, R. Prozorov, and P. C. Canfield, Phys. Rev. B {\bf 87}, 075120 (2013).

\bibitem{Hutchings} M. T. Hutchings 1965 \textit{Solid State Physics: Advances in Research and Applications} \textbf{Vol.16} edited by F. Seitz and B. Turnbull (New York: Academic) p.227.

\bibitem{Stevens} K. W. H. Stevens  \textit{Proc. Phys. Soc.} (London)  \textbf{Sect.A65}, 209 (1952).

\bibitem{Runciman} W. A. Runciman, Phil. Mag. Ser. {\bf 81}, 1075 (1956).

\end{thebibliography}
\end{document}